\documentclass[a4paper]{article}
\usepackage{graphicx}
\usepackage{booktabs}
\usepackage{xcolor}
\usepackage{amssymb}
\usepackage{amsfonts}

\usepackage{lineno}
\usepackage{newtxmath}
\usepackage{natbib}
\usepackage{geometry}
\usepackage{comment}
\usepackage{subfiles} 
\usepackage{academicons}
\definecolor{orcidlogocol}{HTML}{A6CE39}

\newcommand{\lapprox} {\, \lower3pt\hbox{$\sim$}\llap{\raise2pt\hbox{$<$}}\,}
\newcommand{\gapprox} {\, \lower3pt\hbox{$\sim$}\llap{\raise2pt\hbox{$>$}}\,}

\chardef\us=`\_

\begin{document}
\Large
\textbf{AI-FLARES: Artificial Intelligence for the Analysis of Solar Flares Data}

\normalsize

Michele Piana$^{1,2}$, Federico Benvenuto$^1$, Anna Maria Massone$^{1}$, Cristina Campi$^{1}$, Sabrina Guastavino$^1$, Francesco Marchetti$^3$, Paolo Massa$^4$, Emma Perracchione$^5$, Anna Volpara$^1$ \\

\hspace{-0.5cm}$^1$ MIDA, Dipartimento di Matematica, Università di Genova, Genova, Italy  \\
$^2$ Osservatorio Astrofisico di Torino, Istituto Nazionale di Astrofisica, Pino Torinese, Italy \\
$^3$ Dipartimento di Matematica \lq\lq Tullio Levi Civita", Università di Padova, Padova, Italy \\
$^4$ Department of Physics \& Astronomy, Western Kentucky University, Bowling Green, KY 42101, USA \\
$^5$ Dipartimento di Scienze Matematiche \lq\lq Giuseppe Luigi Lagrange", Politecnico di Torino, Torino, Italy

\date{\today}

\begin{center}
   \textbf{Abstract} 
\end{center}

AI-FLARES (Artificial Intelligence for the Analysis of Solar Flares Data) is a research project funded by the Agenzia Spaziale Italiana and by the Istituto Nazionale di Astrofisica within the framework of the ``Attività di Studio per la Comunità Scientifica Nazionale Sole, Sistema Solare ed Esopianeti'' program. The topic addressed by this project was the development and use of computational methods for the analysis of remote sensing space data associated to solar flare emission. This paper overviews the main results obtained by the project, with specific focus on solar flare forecasting, reconstruction of morphologies of the flaring sources, and interpretation of acceleration mechanisms triggered by solar flares. 

$$ $$  
\textbf{key words.} Sun: flares -- Data: EUV, magnetograms, hard X-rays -- Methods: artificial intelligence


\section{Introduction}

Solar flares are the most explosive and energetic events that characterize the active Sun \citep{1988psf..book.....T}. They may extend for more than 10,000 kilometers, release more than $10^{32}$ ergs in less than $100$ seconds, accelerate billion of tons of materials to more than one kilometer per hour, and generate electromagnetic radiation at all wavelengths. From a physical viewpoint, this source of radiation is characterized by very low resistance and very high inductance, so that the rising phase of a flare should last for an (impossibly) long time, which is in contrast with respect to empirical observations. This flare paradox, together with the fact that very little is known about the acceleration and energy release mechanisms within the flaring region, are the reasons why solar flares are still a hot topic in both experimental and theoretical solar plasma physics. Further, from a technological viewpoint, flares are the trigger of space weather \citep{moldwin2022introduction}, i.e., the physical and phenomenological state of natural space environments, which may notably impact the technological assets at earth.

Two general investigation issues concerning solar flares are:
\begin{enumerate}
    \item Are the information contained in the magnetic fields constrained within solar active regions (ARs) and measured by means of magnetograms accurate enough to allow a reliable flare forecasting process?
    \item Which physical mechanisms determine the acceleration of the electrons in coronal plasma, thus triggering the process that leads to the emission of high energy radiation via bremsstrahlung with the ions of the ambient plasma? 
\end{enumerate}

The ``Artificial Intelligence for the analysis of solar FLARES data (AI-FLARES)'' project, funded in 2019 by the Agenzia Spaziale Italiana and the Istituto Nazionale di Astrofisica under the ``Attività di Studio per la comunità scientifica nazionale Sole, Sistema Solare ed Esopianeti'' framework, recognized that these issues can be accomplished by means of two different approaches. From the one hand, magnetohydrodynamics (MHD) equations can be numerically reduced in order to simulate the flaring mechanisms. However, this approach is significantly hampered by the complexity of these partial differential equations and by the limited accuracy of the numerical approximation of their solutions. On the other hand, data-driven approaches can exploit the notable amount of solar, heliophysics, and space weather missions that are currently operating and that can provide an unprecedented amount of multi-modal measurements concerning essentially all possible manifestations of the active Sun. Given this available wealth of data, AI-FLARES focused on the formulation and implementation of computational methods for their interpretation, with applications to the forecasting and modelling of solar flares. Specifically, AI-FLARES developed computational methods for the prediction of the flaring emission and the identification of flare precursors, the reconstruction of flare morphologies for intense eruptive events, and the comprehension of the energy release and acceleration mechanisms for both thermal and non-thermal electrons. The project's objectives were to
\begin{itemize}
    \item Reconstruct the saturated EUV signal in the core region of images of flaring storms recorded by the Atmospheric Imaging Assembly on-board the Solar Dynamics Observatory (SDO/AIA) \citep{lemen2012atmospheric}.
    \item Provide an imaging-spectroscopy picture of the acceleration mechanisms at the base of solar high-energy emissions by exploiting visibilities recorded by both the Reuven Ramaty High Energy Solar Spectroscopy Imager (RHESSI) \citep{hurford2003rhessi} and the Spectrometer/Telescope Imaging X-rays (STIX) on-board Solar Orbiter \citep{2020A&A...642A..15K}.
    \item Design flare forecasting processes and identify the most significant precursors of intense flares by applying machine learning and deep learning algorithms to the Helioseismic and Magnetic Imager on-board SDO (SDO/HMI) magnetograms \citep{scherrer2012helioseismic}. 
\end{itemize}

The methodological inspiration of AI-FLARES relied on an extended interpretation of artificial intelligence, including supervised machine and deep learning, image processing, inverse problems and inverse diffraction theory. From a technological viewpoint, the outcome of this research effort has been a set of computational pipelines for the interpretation of flare-related physics that can be reached at https://github.com/theMIDAgroup/AI-FLARES.

The plan of the paper is as follows. Section 2 describes AI-FLARES results concerned with flare forecasting. Section 3 and Section 4 focus on image processing and reconstruction at EUV and hard X-ray wavelengths, respectively. Our conclusions are offered in Section 5.

\section{AI-FLARES and solar flare forecasting from magnetograms}

In the last decade a notable amount of scientific literature has been illustrating the results of the application of data-driven AI-based approaches to flare forecasting \citep{2018SpWea..16....2C,2018ApJ...856....7H,2018SoPh..293...28F,2019ApJ...881..101L,2021JSWSC..11...39G,2021A&C....3500468R,2021EP&S...73...64N,2022ApJ...935...45S}. Most of these studies describes the action of either machine learning tools that process features extracted from magnetograms, or deep neural networks that directly take as input full disk images of solar active regions (ARs). On the one hand, a possible objective of feature-based machine learning is to identify which AR descriptors mostly impact the forecasting process. On the other hand, deep learning aims at improving the predictive power hidden within the space data by means of black-box approaches that take as input images or videos of AR magnetograms and provide as output a binary classification based on features that are automatically computed by the neural networks. 

\begin{figure*}[t!]
\resizebox{\hsize}{!}{\includegraphics[clip=true]{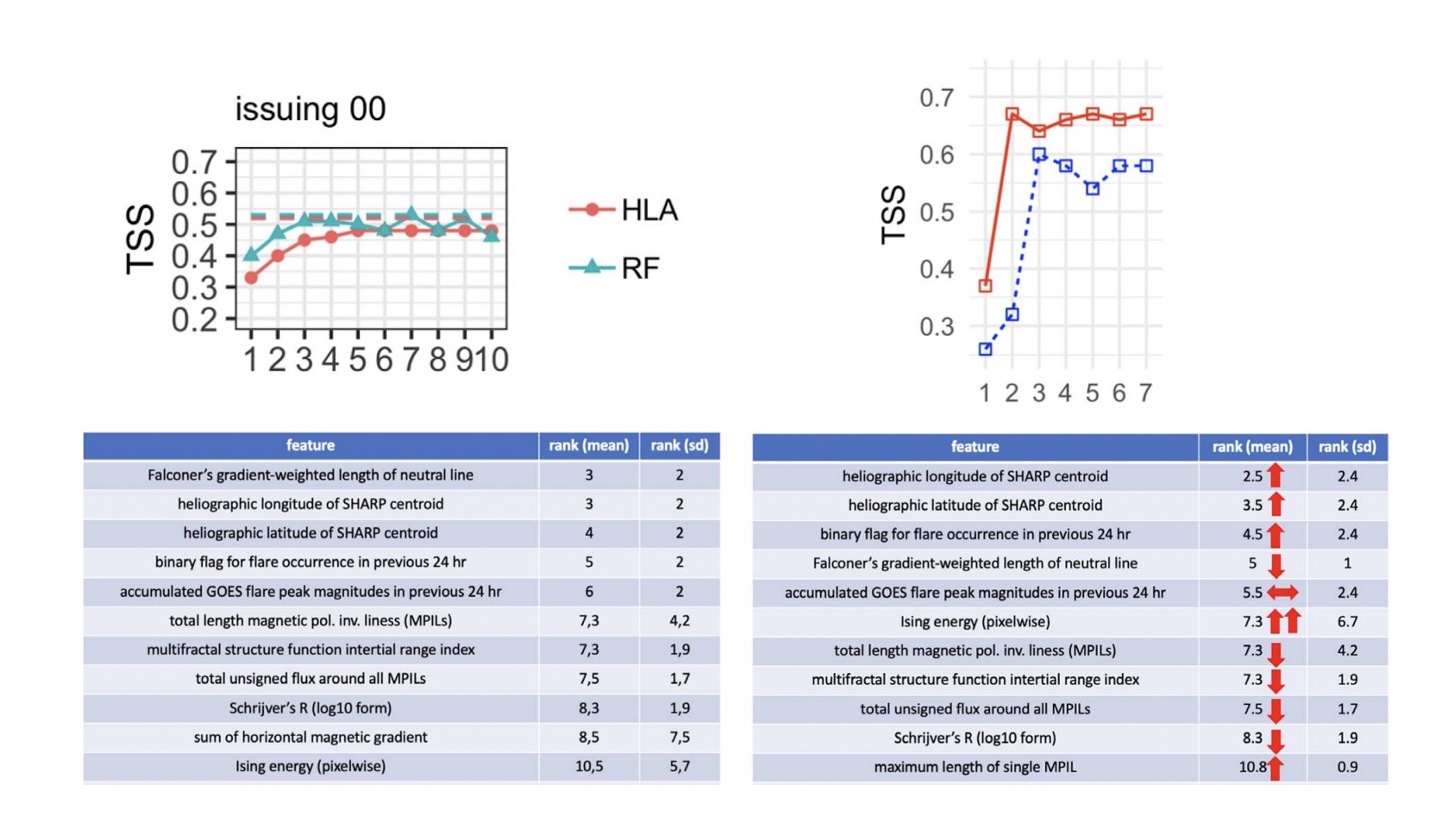}}
\caption{\footnotesize
{\bf Feature-based machine learning for flare forecasting.}
Top left panel: very few descriptors (x-axis) are sufficient to achieve high values of the True Skill Statistic (TSS) score (y-axis) with two machine learning methods (hybrid LASSO, HLA and Random Forest, RF). Bottom panel: feature-ranking applied to machine learning outcomes show that the Ising energy significantly increases its rank when the AR producing the flaring event is included in the training set. Top right panel: the TSS notably increases when the topological feature introduced in \cite{2021ApJ...915...38C} is added as first descriptor (red solid line), with respect to the case wehm the feature is not used in the training set (blue dashed line).
}
\label{fig:figure-1}
\end{figure*}

AI-FLARES addressed the first issue by means of sparsity-enhancing regularization methods applied to
the SDO/HMI archive in specific but large time
ranges \citep{2018ApJ...853...90B,2018SoPh..293...28F,massone2018machine,piana2018flarecast,2019ApJ...883..150C,2020ApJ...904L...7B,2021ApJ...915...38C}. In a typical pipeline of analysis, the HMI magnetograms have been grouped into four subsets belonging to the four issuing times 00:00, 06:00, 12:00, and 18:00 UT. For each AR we used the $171$ features extracted by means of the algorithms implemented within the FLARECAST Horizon 2020 project \citep{2021JSWSC..11...39G}. For each subset, i.e., for each issuing time, we generated the training, validation, and test sets and, for each sample in the training set, the annotation was performed providing label "1" to an event occurred within $24$ hours from the issuing time. By applying AI-FLARES machine learning algorithms to these data we were able to prove that (see Figure \ref{fig:figure-1}):
\begin{itemize}
\item Very few AR descriptors are really effective in the forecasting process and these descriptors are very robust, independently of the regularization method used and of other experimental aspects like the issuing times considered in the training set \citep{2019ApJ...883..150C}.
\item The Ising energy seems to systematically play a notable predictive role, specifically in the case of the forecasting of particularly intense flaring storms \citep{2020ApJ...904L...7B}.
\item The computation of innovative topological descriptors can represent a way to improve the skill scores associated to feature-based machine learning algorithms \citep{2021ApJ...915...38C}.
\end{itemize}

\begin{figure*}[t!]
\resizebox{\hsize}{!}{\includegraphics[clip=true]{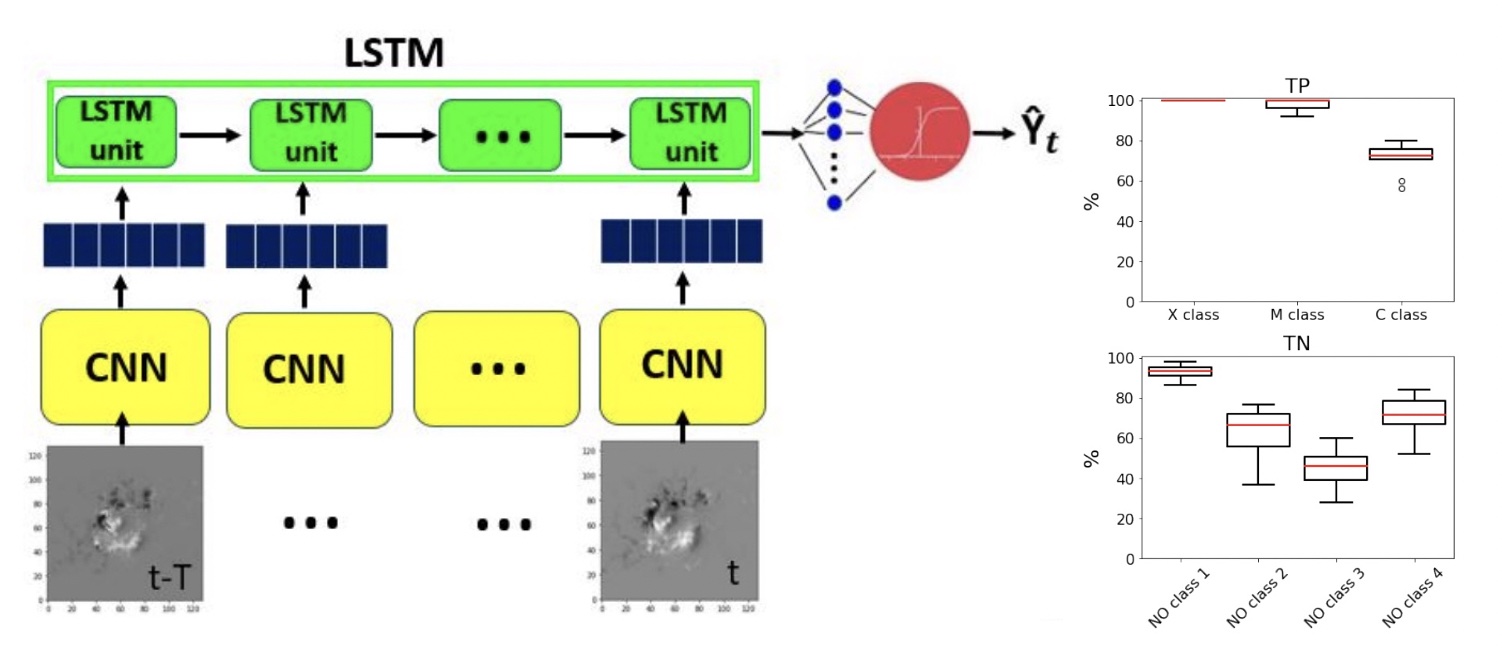}}
\caption{{\bf Deep learing for flare forecasting.} Left panel: the AI-FLARES neural network is made of a Long Short-Term Memory (LSTM) network fed by the outcomes of several Convolutional Neural Networks (CNNs). Right panels: the rates of true positives and true negatives are significantly high thanks to the use of video-based deep learning.
}
\label{fig:figure-2}
\end{figure*}

The second AI-FLARES perspective on flare forecasting was the development of deep learning networks able to provide a notable prediction improvement in the difficult game of space weather prediction. In this context, the main result of AI-FLARES (see Figure \ref{fig:figure-2}) has been the implementation of a pipeline that, for the first time, utilizes videos of HMI frames as input data and that, again for the first time, accounts for an appropriate balancing of different data sample types in the training and validation phases of the neural network \citep{2022A&A...662A.105G}. We implemented the AI-FLARES Long-term Recurrent Convolutional Network (LRCN) and validated it against the Near Realtime Space Weather HMI Archive Patch (SHARP) data products associated with the line-of-sight components in the time range between 2012 September 14 and 2017 September 30. More specifically, we used 24-h-long videos made of 40 SHARP images of an AR, with 36 minutes cadence. These videos have been categorized as C-, M-, and X-class flares and also according to four different null-events classes. We generated a training set, a validation set and a test set, and we used data augmentation to increase their cardinality. We repeated this set generation process ten times in order to create ten random realizations of these three sets. The algorithm for set generation was inspired to two strategic principles: proportionality, i.e., use of the same rates of samples for each category; and parsimony, i.e., use of as few ARs as possible so that samples belonging to the same AR fall into the same set. As shown in Figure \ref{fig:figure-2}, the true positive rates provided by the deep network are significantly high (in particular, both M and X classes are more distinguishable from ARs associated with NO-class),  and in all cases the standard deviations are nicely small.

As a final comment, we point out that AI-FLARES provided contributions also to the methodological field related to machine and deep learning research. In particular, during this project two theoretical ideas have been conceived and formulated. The first one is about the use of probabilistic score-oriented loss functions in the training phase for neural networks \citep{2022PatRe.13208913M}; the second one is about the use of value-weighted skill scores for the performance assessment of both machine and deep learning \citep{9813503,guastavino2023operational}. These two methodological tools have been utilized in most networks designed for the flare forecasting approaches developed within AI-FLARES.

\section{AI-FLARES and EUV image de-saturation}
EUV measurements recorded in correspondence with intense solar flares are almost systematically affected by saturation. After the launch of SDO/AIA, the desaturation of EUV images has become a big data issue, since AIA has been providing more the $10^5$ frames per year since February 2010 \citep{2014ApJ...793L..23S,2015InvPr..31i5006T,2015A&C....13..117S}. 

\begin{figure*}[t!]
\resizebox{\hsize}{!}{\includegraphics[clip=true]{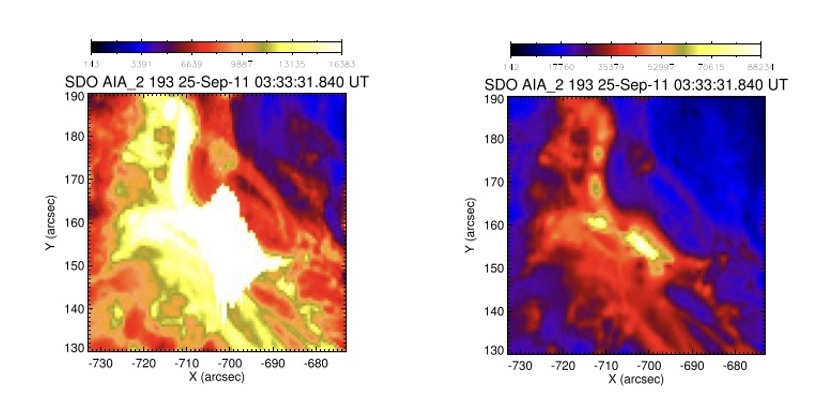}}
\caption{{\bf Desaturation of EUV maps.} Left panel: an intense solar flare saturates an extended region of an image recorded by SDO/AIA. Right panel: AI-FLARES desaturation algorithm is able to restore the signal in the core of the flaring source.
}
\label{fig:fig-3}
\end{figure*}

AI-FLARES contribution to image processing at the solar EUV regime is represented by the formulation and implementation of the Sparsity-Enhancing DESAT (SE-DESAT) method, a novel computational approach for the analysis of SDO/AIA saturated images able to recover the signal in the primary saturation region in a rapid fashion without using any other information but the one contained in the image itself \citep{2019ApJ...882..109G}. SE-DESAT is a modification of a previous algorithm developed in our group, named DESAT \citep{2015AC....13..117S}. As for DESAT, also in SE-DESAT the input data are represented by the diffraction fringes and therefore this is again an inverse diffraction algorithm. However, unlike for DESAT, this new approach realizes segmentation between the primary saturation region and blooming, background estimation and desaturation in the primary saturation region at the same time, without the need of any a priori information on the image background. Further, an adaptive version of SE-DESAT (adaptive SE-DESAT) introduced weights depending on the shape of the saturated region \citep{2021InvPr..37a5010G}. An example of how SE-DESAT performs is described in Figure \ref{fig:fig-3}, in the case of EUV images recorded by AIA on September 25 2011, in the $193$ $\AA$ bandwidth, at time point 03:33:31 UT.

We point out that the availability of this desaturation pipeline has currently two important consequences. First, scientists working in the STIX community are used to integrate the hard X-ray information contained in the STIX data with the EUV information contained in the AIA data, and to this aim de-saturated AIA images are needed \citep{2022SoPh..297...93M}. Second, AI-FLARES people are currently involved in a NASA project for nowcasting solar flares starting from AIA information \citep{2022FrASS...940099M}. Once again, image desaturation will be a crucial pre-processing step for the realization of a prediction approach that will apply machine learning to de-saturated EUV maps.

\section{AI-FLARES and hard X-ray imaging spectroscopy}

\begin{figure*}[t!]
\resizebox{\hsize}{!}{\includegraphics[clip=true]{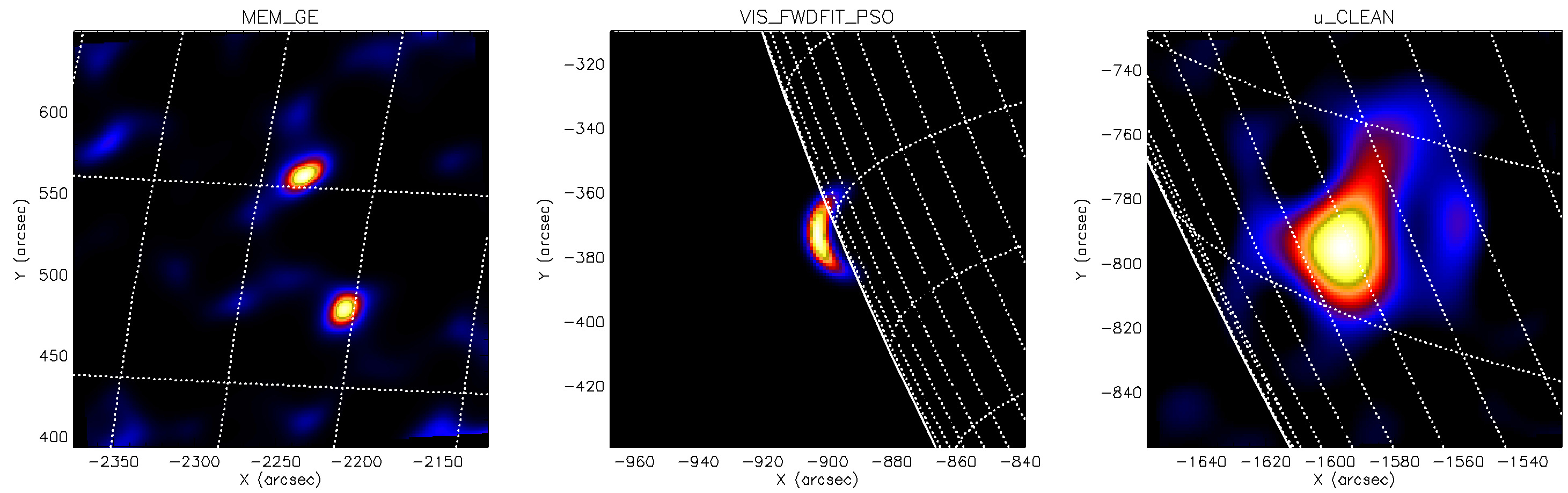}}
\caption{{\bf Image reconstruction from hard X-ray visibilities.} A constrained maximum entropy method, a forward-fit algorithm based on Particle Swarm Optimization (PSO), and an automated version of CLEAN deconvolution metod provided the reconstructions in the left, middle, and right panels, respectively.}
\label{fig:fig-4}
\end{figure*}

\begin{figure*}[t!]
\resizebox{\hsize}{!}{\includegraphics[clip=true]{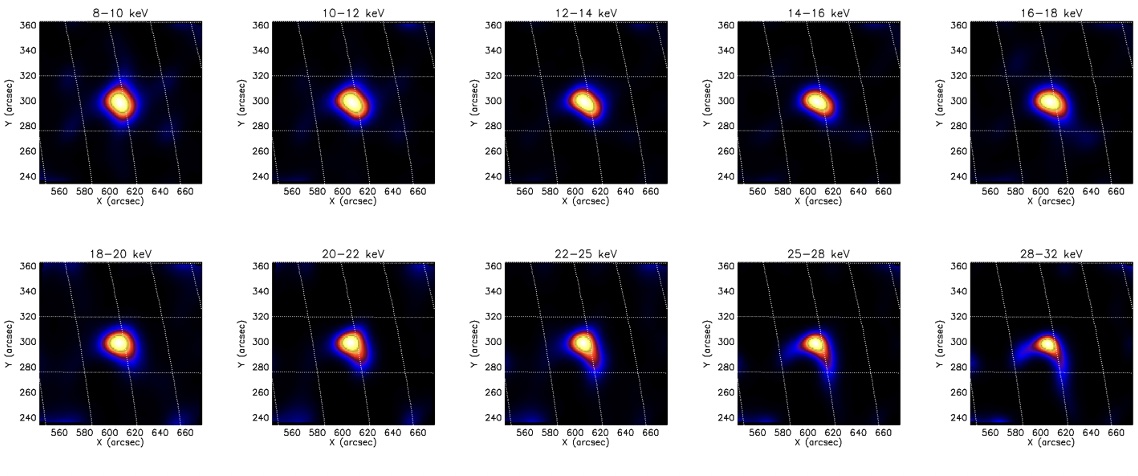}}
\caption{{\bf Regularized imaging spectroscopy from hard X-ray visibilities.} These panels represent electron flux maps at different electron energies (regularization introduced a smoothing constraint across the energy direction.)
}
\label{fig:fig-5}
\end{figure*}

AI-FLARES results concerned with hard X-ray imaging spectroscopy \citep{piana2022hard} have been of three kinds. We formulated the mathematical model for the image formation process of STIX visibilities and contributed to the calibration process for all thirty STIX collimators devoted to imaging \citep{2019A&A...624A.130M,2020A&A...642A..15K,2023SoPh..298..114M}. We then developed several image reconstruction methods able to represent in the image space the information contained in the hard X-ray visibilities recorded by either RHESSI or STIX. Specifically, we have formulated, implemented and validated (see Figure \ref{fig:fig-4}):
\begin{itemize}
\item A maximum entropy method, in which the solution is constrained to have positive entries and total flux equal to an a priori estimate \citep{2020ApJ...894...46M}.
\item An interpolation/extrapolation method based on feature augmentation and on the use of Variably Scaled Kernels \citep{2021ApJ...919..133P,2021InvPr..37j5001P} that allows the implementation of an automated version of CLEAN deconvolution \citep{2023ApJS..268...68P}.
\item A parametric imaging method that works for both a partial information on the visibility set, i.e., when just the visibility amplitudes are provided by the instrument \citep{2021A&A...656A..25M} and when visibilities are recorded by fully calibrated collimators \citep{2022A&A...668A.145V} (the algorithm relies on Particle Swarm Optimization).
\end{itemize}
Finally, and probably more importantly, as part of the RHESSI legacy \citep{2007ApJ...665..846P,prato2009regularized}, we formulated and implemented a regularization method that is able to reconstruct maps whose pixel content is proportional to the average flux of the electrons accelerated along the magnetic field lines from the corona down to the chromosphere. The nicest aspect of this methodological approach (see Figure \ref{fig:fig-5}) is that it is able to provide electron maps that are constrained to vary in a smooth way along the spectral direction and that can be projected back to the photon domain to produce photon maps that are in turn regularized across energy. This approach may represent an important step to a full interpretation of STIX data within the framework of imaging spectroscopy, and may provide a crucial tool for the understanding of electron acceleration mechanisms during solar flares \citep{2023arXiv231107148V}.

\section{Conclusions}
This paper has reviewed the main results obtained within the framework of the AI-FLARES project. These results involve the whole workflow concerning the interpretation of solar flare data, starting from the prediction of flaring events using magnetograms, through the restoration of morphological aspects of the flaring sources in saturated high-resolution EUV maps, to the realization of an innovative imaging spectroscopy approach in the case of hard X-ray visibilities. 

The legacy of AI-FLARES in the current activity of our research group is two-fold. On the one hand, we are developing neural networks for space weather forecasting that encode physical information in the loss function utilized for the training \citep{2023ApJ...954..151G}. On the other hand, we are studying how regularized electron maps can be used to obtain quantitative information about the effectiveness of the electron acceleration mechanisms triggered by magnetic reconnection high in the solar corona \citep{2023arXiv231107148V}.

\section*{Acknowledgements}
All authors strongly acknowledge the financial support of the "Accordo ASI-INAF AI-FLARES n. 2018-16-HH.O" grant.

\bibliographystyle{aa}
\bibliography{bibliography}

\end{document}